\documentclass[aps,showpacs,twocolumn]{revtex4}
\usepackage{graphicx}
\usepackage{bm}
\usepackage[squaren]{SIunits}

\begin{document}
\title{Magnetoelectric studies on polycrystalline FeVO$_{4}$}
\author{L. Zhao$^1$, M.P.Y. Wu$^2$, K.W. Yeh$^1$, M.K. Wu$^1$}
\email{mkwu@phys.sinica.edu.tw}
\affiliation{$^1$Institute of Physics, Academia Sinica, Taipei 11529, Taiwan\\
$^2$Electrical Engineering Department, University of California, Los Angeles. CA 90024, USA}
\date{\today}

\begin{abstract}
We report the magnetic, dielectric and ferroelectric properties of  polycrystalline iron vanadate(FeVO$_{4}$), which has been recently found to exhibit multiferroicity in low temperature  with noncollinear magnetic orderings. The influence of external magnetic field up to 9T on these properties is systematically investigated. Besides the suppressing effect on the original ferroelectric transition, the strong magnetic field seems to induce a secondary ferroelectric transition at a slight lower temperature. And the corresponding magnetization measurement reveals a field-induced metamagnetic transition at low temperatures. Our results will help to clarify the complex magnetic structure and microscopic mechanism of multiferroicity in  FeVO$_{4}$.
\end{abstract}
\maketitle

\section{Introduction}

Stimulated by the recent discovery of strong coupling of magnetism and ferroelectricity in some frustrated manganites as TbMnO$_{3}$  and TbMn$_{2}$O$_{5}$ \cite{TbMnO3,TbMn2O5}, the research on multiferroics has attracted worldwide attention. In these materials, both magnetic and ferroelectric orders coexist. Although this kind of coexistence has been discovered decades ago, most known multiferroic materials, e.g. BiFeO3, show  very weak magnetoelectric coupling because their ferroelectricity and magnetism come from different ions in the unit cell, with partially filled $d$ shells and empty $d$ shells respectively. These two orders, therefore, tend to be mutually exclusive and interact weakly with each other\cite{Fiebig}. On the contrary, the ferroelectricity found in  these frustrated manganites is of magnetic origin, i.e., induced by complex spin configuration. The spontaneous electric polarization occurs in special magnetically ordered states which break the inversion symmetry, and the electric polarization can be switched and  reversed via changing the magnetic states by external magnetic field. These fascinating phenomena are of great importance both for the fundamental physics and potential technological application \cite{CheongNM, LiuJM}.

There are several reported types of magnetic orderings which can break the space inversion invariance and produce spontaneous electric polarization. For example, the simple inequivalent nearest-neighbor exchange striction in the commensurate collinear $\uparrow\uparrow\downarrow\downarrow$ spin chains can shift ions away from centrosymmetric positions and induce the electric polarization along the spin chain. But most of  the realistic multiferroic frustrated magnets are of much more complex spin configurations, which are usually noncollinear  and incommensurate at low temperatures. One of the commonly accepted microscopic mechanism comes down to the inverse Dzyaloshinskii-Moriya(DM) interaction, which was first proposed as an antisymmetric relativistic correction to the superexchange coupling\cite{DM}. It can also be expressed as the equivalent spin current model by Katsura, Nagaosa and Balatsky(KNB)\cite{Nagaosa}.  According to their theoretic analysis, the microscopic polarization induced by neighboring spins can be formulated as  ${\bf P}_{ij}=A\hat{\bf {e}}_{ij}\times({\bm S}_i\times{\bm S}_j)$, where $A$ is the coupling constant determined by the spin-orbit coupling and exchange interactions, and $\hat{\bf {e}}_{ij}$  the unit vector connecting site $i$ and $j$. It is also consistent with the corresponding phenomenological symmetry analysis\cite{Mostovoy}, and has worked well in the helimagnetic TbMnO$_{3}$ and other multiferroics. But at present the accurate prediction of the magnitude still lacks since much more complicated factors in real systems must be considered\cite{Dagotto, LiuJM}.

On the other hand, according to the known microscopic mechanism of magnetoelectric coupling,  the polarization and dielectric properties of multiferroic compounds can also help to explore the complex magnetic structures, which traditionally have to be characterized by the much more difficult and expensive experimental techniques such as neutron diffraction \cite{Lawes}.

\begin{figure}
\includegraphics[width=0.35\textwidth]{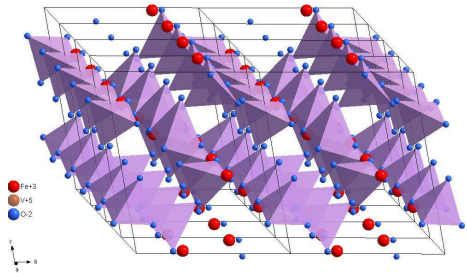}
\caption{
Schematic crystal structure of FeVO$_{4}$
}
\label{fig.1}
\end{figure}

At present, our particular interest is on the recently discovered multiferroicity in iron vanadate(FeVO$_{4}$). The crystal structure  of FeVO$_{4}$ has been investigated in the early 1970s\cite{1971, 1972}.  FeVO$_{4}$ have a triclinic crystal structure (Ref:PDF number 71-1592 space group $P1\bar{1}$, $a$=6.719\AA , $b$=8.060\AA , $c$=9.254\AA, $\alpha$=96.4$^\circ$,$\beta$=106.8$^\circ$, $\gamma$=101.5$^\circ$), which is shown in Fig.~\ref{fig.1}.  The magnetism of FeVO$_{4}$ comes from the Fe$^{3+}$ ions at three different sites, which is orbitally quenched and in the high spin $S=5/2$ state. The Fe$^{3+}$ ions form a peculiar chain-like structures, well separated by non-magnetic VO$_{4}$ tetrahedrons.  The complex Fe-O-Fe superexchange (SE) and  Fe-O-O-Fe super-super-exchange(SSE) interactions between Fe$^{3+}$ ions lead to the magnetic frustration in FeVO$_{4}$,  and therefore complex magnetic behavior.

 Disclosed by the recent measurements of susceptibility and heat capacity \cite{2008, Kundys1, Kundys2, Lawes1, Lawes2},  FeVO$_{4}$ undergos two successive magnetic transitions at  T$_{{\rm N1}}\simeq$22K and T$_{{\rm N2}}\simeq$15K respectively. The preliminary neutron diffraction experiment (in zero field) reveals a spiral spin structure at lower temperature (T$<$T$_{{\rm N2}}$), while a collinear antiferromagnetic(AFM) ordering in the intermediate state(T$_{{\rm N2}}$$<$T$<$T$_{{\rm N1}}$)\cite{Kundys2}. The two kinds of magnetic structures are both incommensurate but the ferroelectricity is found only concomitant with the  non-collinear magnetic order at low temperature. The physical nature of its multiferroicity, however,  remains elusive. The detailed magnetic structure and its further evolution with external field deserve further in-depth research. Till now, there is only a few reports on the multiferroic FeVO$_{4}$. Therefore, in this paper, we carried out the systematically dielectric, ferroelectric and magnetic measurements on  polycrystalline FeVO$_{4}$ at low temperatures. Out work will help to determine the complex magnetic structures and further unravel the  microscopic mechanism of the multiferroicity in FeVO$_{4}$.

\section{Experimental}
The polycrystalline FeVO$_{4}$  are prepared by conventional solid-state reaction using high-purity Fe$_2$O$_3$ and V$_2$O$_5$ as staring materials. The detailed procedures have been described elsewhere\cite{1971, 1972}.  It is notable that V$_2$O$_5$ is  volatile and reacts with alumina crucible at high temperature, which leads to the heavy contamination of the samples. The use of alumina crucibles should therefore be avoided, and instead, a platinum crucible with a cover is recommended in the sinter procedure.

Magnetic properties were measured on a SQUID magnetometer(MPMS-5S, Quantum Design) and a 9T-Physical Properties Measurement System (PPMS) with a magnetization option. To measure dielectric properties of  FeVO$_{4}$, the polycrystalline bulk is polished to thin plates with thickness of $0.1-0.2$mm. We use silver epoxy attached to both sides as electrodes to form a parallel plate capacitor whose capacitance is proportional to the dielectric constant($\epsilon_r$).  The samples are glued on the cryogenic stage of our homemade probe, and connected to a high-precision capacitance meter. The main  sources of  error as residual impedance in the whole circuit are carefully considered and compensated. Our measure system has been tested with the standard commercial  capacitors(0.5-10pF), which is close to our samples.
For temperature-dependent dielectric measurements, the sample's capacitance  was measured over a range of frequencies with an excitation of 1 V, while the temperature was swept at a very slow warming or cooling rate(0.1K-0.5K/min) to avoid the thermal  inhomogeneity and lagging feedback of the Cernox thermometer. We also tried different excitations(50mV to 1V), sweeping rates and so on. No apparent difference was found in the different conditions.

The electric polarization ($P$) is obtained from the integration of the measured pyroelectric current. We first poled the sample during the cooling process with applied electric field of over 400kV/m, then removed the electric field and shorted the sample at lowest temperature for at least 30 minutes to discharge the free charge carriers. The pyroelectric current was measured with a fixed warming rate of 3K/min.

\begin{figure}
\includegraphics[width=0.35\textwidth]{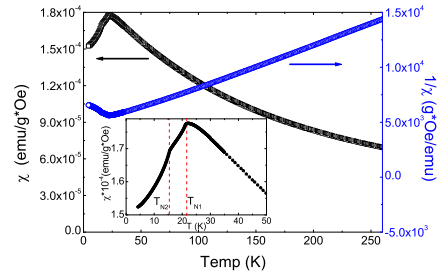}
\caption{
the temperature-dependent magnetic susceptibility and the corresponding inverse susceptibility of FeVO$_{4}$ measured in a magnetic field of 1000Oe. The susceptibility at low temperatures is also shown in the inset with two vertical dashed lines marking the magnetic transitions at T$_{{\rm N1}}$ and T$_{{\rm N2}}$.
}
\label{fig.2}
\end{figure}

\section{Results and Discussions}
Fig.~\ref{fig.2} shows the temperature dependence of magnetic susceptibility($\chi$) and corresponding inverse one($\chi^{-1}$) measured in the field of H=1000Oe. At high temperature(T$>$30K), a typical Curie-Weiss paramagnetic behavior is observed, i.e., $\chi=C/(T-\theta)$. The effective magnetic moment($\mu_{eff}$) in the paramagnetic phase is about 5.96$\mu_{B}$, which is deduced from the fitted Curie constant $C$. It is very close to 5.916$\mu_{B}$, the expected value for $S=5/2$ Fe$^{3+}$ ions in the high spin state.  The fitted $\theta$=-112K suggests the main antiferromagnetic interaction existing between Fe$^{3+}$ ions.

As temperature decreases, two successive magnetization anomalies are observed at T$_{{\rm N1}}$=21.7K and T$_{{\rm N2}}$=15.4K. This is in good agreement with the previous experiments \cite{1971,1972,Kundys1,Lawes1}, suggesting two different magnetic ordering states at low temperature, T$_{{\rm N2}}$$<$T$<$T$_{{\rm N1}}$ and T$<$T$_{{\rm N2}}$ respectively.  As disclosed in the neutron diffraction experiment\cite{Kundys2}, the helical magnetic order exists as T$<$T$_{{\rm N2}}$ while the collinear AFM spin density wave exists in the intermediate state(T$_{{\rm N2}}$$<$T$<$T$_{{\rm N1}}$).  It is also noted that the transition temperatures reported by He et al \cite{2008}(20K and 13K respectively), are quite lower, possibly due to the Al contamination from the alumina crucibles used in the crystal growth process.

\begin{figure}
\includegraphics[width=0.35\textwidth]{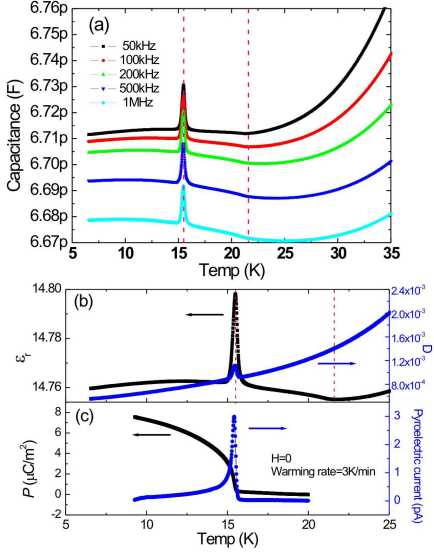}
\caption{
(a) the  temperature dependence of the capacitance of a typical FeVO$_{4}$ sample measured at different testing frequencies in zero field.
(b) the dielectric constant($\epsilon_r$) and corresponding tan loss(D)measured at freq=100kHz.
(c) pyroelectric current measured at a warming rate 3K/min and corresponding electric polarization(P)
}
\label{fig.3}
\end{figure}

Fig.~\ref{fig.3}(a) shows our raw data of the sample capacitance in zero field, which is proportional to the dielectric constant of FeVO$_{4}$. The most remarkable feature is the sharp peaks of dielectric anomaly at T$_{{\rm N2}}$, which is clearly observed for all our testing frequencies, indicating the high quality of our sample. Evidently, these dielectric peaks are associated with the emerging of ferroelectric order. Around T$_{{\rm N1}}$, there is a slight but not non-negligible discontinuity in $\epsilon_r$(T)(see (b) for details). A small excess increase of $\epsilon_r$ occurs as T$<$T$_{{\rm N1}}$ independent of the testing frequency, which excluds the extrinsic factors as interface effect. The slight anomaly at T$_{{\rm N1}}$ is also consistent the recent observation on  FeVO$_{4}$ thin films by Dixit et al\cite{Lawes2}.  For convenience, we adopt only the data measured at 100kHz in our following discussions.

Further evidence of the ferroelectric transition at T$_{{\rm N2}}$ is shown in Fig.~\ref{fig.3}(b). The sharp peak in the dielectric loss at T$_{{\rm N2}}$  coincides with the peak in  $\epsilon_r$(T), which is a typical feature of proper ferroelectric transition. But no peak or anomaly of dielectric loss appears around T$_{{\rm N1}}$. So we confirms that the ferroelectricity is only concomitant with the noncollinear magnetic state in low temperature($<$T$_{{\rm N2}}$), consistent with the inverse DM mechanism.

We also measured the corresponding electric polarization by integrating the pyroelectric current. $P$ is about 7.3$\mu$C/m$^{2}$ as T=10K, in agreement with previous measurements\cite{Kundys1, Lawes1, Lawes2}. The $P$ can be reversed by an opposite poling electric fields (not shown here).

\begin{figure}
\includegraphics[width=0.35\textwidth]{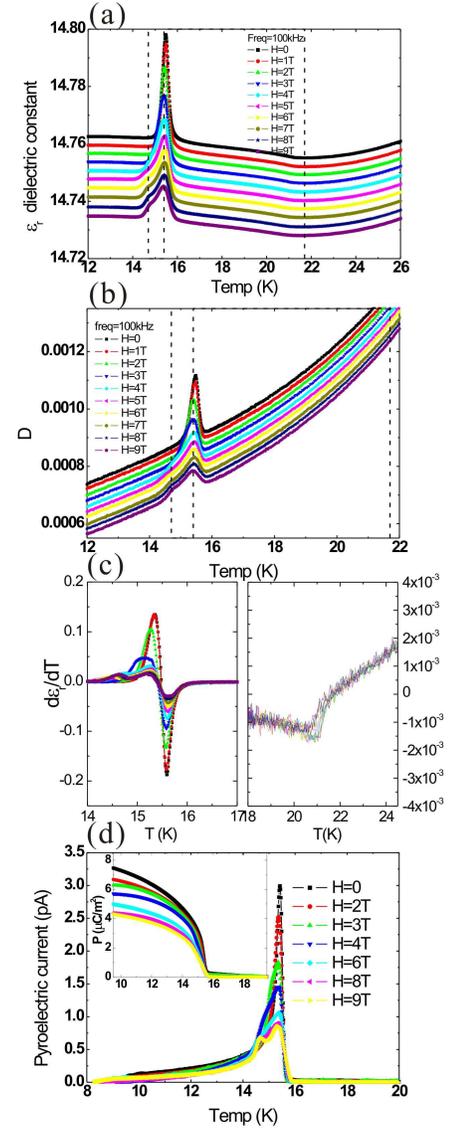}
\caption{
(a) and (b) are the temperature dependence of dielectric constants($\epsilon_r$) and corresponding dielectric loss(D) measured in different field(H=0-9T). the data are shifted vertically for clarification except the H=0 case.  The vertical dashed lines mark the different transition temperatures, T$_{{\rm N1}}$(21.7K), T$_{{\rm N2}}$(15.4K)  and T$_{{\rm N3}}$(14.7K) respectively.
(c) is the temperature derivative of dielectric constants, d$\epsilon_r$/dT.
(d)  the pyroelectric currents measured in different H from 0 to 9T with the same warming rate of 3K/min. The corresponding electric polarization ($P$) is shown in the inset .
}
\label{fig.4}
\end{figure}

To further disclosed the magnetoelectric coupling and the possible effect of external magnetic field on the complex spin structures, we measured the $\epsilon_r$(T) in different magnetic fields. In our experiments, there is no discernable difference between the two H$\bot$E and H$\parallel$E measuring configurations, since the intrinsic anisotropy in FeVO$_{4}$ is canceled out in our polycrystalline bulk.

As shown in Fig.~\ref{fig.4}(a),  the slight discontinuous increase in $\epsilon_r$(T) around T$_{{\rm N1}}$, which is associated with the first magnetic transition, remain almost unchanged by magnetic field up to 9T. To demonstrate the fine variance in  $\epsilon_r$(T) induced by H,, the corresponding temperature derivative of $\epsilon_r$(T) is plotted in Fig.~\ref{fig.4}(c).

The most remarkable feature induced by magnetic field occurs around the second magnetic transition at T$_{{\rm N2}}$($\sim$15.4K). The sharp peak in $\epsilon_r$(T) in zero field is gradually suppressed by increasing magnetic fields. This behavior also occurs in dielectric loss(Fig.~\ref{fig.4}(b)), indicating that magnetic field suppresses ferroelectricity. However, the corresponding transition temperature, which is defined by the maximum of $\epsilon_r$(T) peak, remains unchanged even in the the highest field(H=9T).

Another significant characteristic is the emergence of a new  dielectric hump around the 14.7K (denoted as T$_{{\rm N3}}$, which is slightly lower than T$_{{\rm N2}}$)  as H is strong enough ($>$3T).  As H grows further, the emerging dielectric hump grows gradually, while  the  peak at T$_{{\rm N2}}$ are further suppressed.  The similar behavior is also observed in D(T)(Fig.~\ref{fig.4}(b)).  Our dielectric measurement has been repeated several times and also on other   FeVO$_{4}$ samples with different sizes to exclude possible extrinsic effects such as space charge and inhomogeneity in ceramics.

All the present results indicate a possible new field-induced new transition at T$_{{\rm N3}}$, which is also a ferroelectric one (deduced from the simultaneous anomalies in $\epsilon_r$ and D independent of the testing frequencies).  To further confirm the nature of this possible transition,  we also carried out the pyroelectric measurements in different H. The pyroelectric current measured at a fixed warming rate (shown in Fig.~\ref{fig.4}(d)) is proportional to the temperature derivative of electric polarization ($dP/dT$) and better reveals the field-induced fine change than the $P$(T) curve itself (see the inset). The  pyroelectric current peak in zero field is very sharp, which marks the narrow ferroelectric transition at T$_{{\rm N2}}$. The peak is suppressed greatly by increasing H. However this suppression by magnetic fields seems to saturate as H$\geq$8T.  When H$>$3T, a secondary peak of the pyroelectric current emerges and develops with growing H, indicating the new additional ferroelectric polarization induced by external field around T$_{{\rm N3}}$. The total $P$ includes the different contributions from two transitions, which are hard to be separated, as shown in the inset. To our knowledge, it may be the first evidence that there exist a field-induced new multiferroic transition in the FeVO$_{4}$ system, which is owed to our accurate and elaborate measurements.

\begin{figure}
\includegraphics[width=0.35\textwidth]{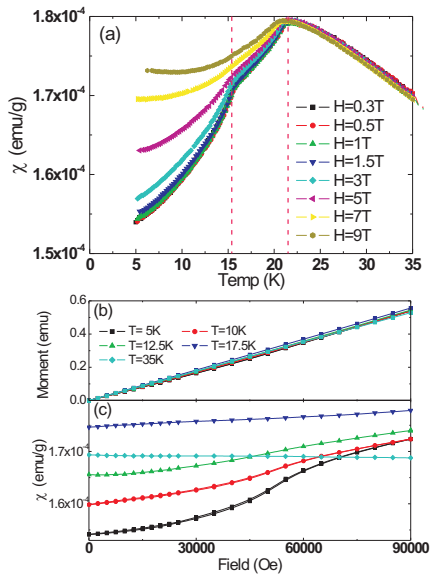}
\caption{
(a) the temperature-dependent magnetic susceptibility of FeVO$_{4}$ measured in different
fields(0.3-9T) on a 9T PPMS. (b) The magnetic field dependence of magnetization of FeVO$_{4}$ measured at
several temperatures(T=5K, 10K, 12.5K, 17.5K, 35K) from 0 to 9T. Corresponding magnetic susceptibilities are also plotted in (c).
}
\label{fig.5}
\end{figure}

Considering the magnetic origin of the ferroelectricity in FeVO$_{4}$, the new multiferroic transition must be closely related the change of spin structures in strong fields. So we performed the magnetization measurements in different fields (H=0.3-9T) and the results are shown in Fig.~\ref{fig.5}(a). The first magnetization anomaly at T$_{{\rm N1}}$ is evident  in all H up to 9T with negligible temperature shift, consistent with the corresponding dielectric measurements(see Fig.~\ref{fig.4}(a)and (c)). The second magnetization anomaly at T$_{{\rm N2}}$ can be clearly discerned only in low fields(H$<$5T),  then gradually fades away with growing up-turn background as H increase further, and finally become indistinguishable in H=9T.

To further explore the fine change of the low-temperature magnetic states in different H, we measured the field-dependent magnetization, which is shown in Fig.~\ref{fig.5}(b). The deviation from linearity in M-H curves is small.  To manifest this deviation, we plotted the corresponding field-dependent susceptibility ($\chi$(H)) in Fig.~\ref{fig.5}(c).

In the high temperature paramagnetic state($>$T$_{{\rm N1}}$), as T=35K, the quite linearity of the M-H curve and nearly constant $\chi$(H) is observed up to 9T.  In the intermediate state between T$_{{\rm N1}}$ and T$_{{\rm N2}}$(T=17.5K), the $\chi$(H) deviates slightly from the constant and grows with increasing H. The increase in $\chi$ come possibly from the partial suppression of the  antiferromagnetic ordering by external magnetic field.

As cooled further down to T$=$ 10 and 5 K (both well below T$_{{\rm N3}}$), In zero field, FeVO$_{4}$ is dominated by the incommensurate helicodal  magnetic ordering, which has been disclosed in recent neutron diffraction experiment\cite{Kundys2}.
The nonlinearity in M-H curves becomes much pronounced as H$>$ 3T. Especially, a broad kinking appears in $\chi$(H) between 3T and 6T. In the high field limit, the  $\chi$(H) at 5K and 10K seems to merge together, suggesting a new common stable magnetic ground state.
Our data indicates a field-induced metamagnetic transition at T$<$T$_{{\rm N2}}$, although its nature is still unclear.
According to our observed ferroelectric behavior and the  the inverse DM mechanism, we conjectured that the new field-induced magnetic phase must be also noncollinear.
The neutron diffraction measurements on FeVO$_{4}$ in strong fields is greatly expected in future to disclose the new magnetic phase in detail.

By the way, the multiferroicity is usually highly anisotropic, while only bulk average properties can be acquired on our polycrystalline samples. So the growth of large-sized single crystals is necessary to explore the anisotropic intrinsic dielectric response and electric polarization to unravel the complex magnetic orderings via magnetoelectric coupling in multiferroic FeVO$_{4}$ \cite{Lawes}.  The effect of the partial substitution of Fe$^{3+}$ by other transition metal ions is also our work under way.

\section{Summary}
We  have investigated systematically the dielectric and ferroelectric properties of polycrystalline FeVO$_{4}$ in external magnetic field as well as the corresponding magnetization. The two successive magnetic transitions occur at low temperatures(T$_{{\rm N1}}$ and T$_{{\rm N2}}$ respectively). The multiferroicity emerging below T$_{{\rm N2}}$  is confirmed and found to be strongly suppressed with increasing external magnetic field. At the same time, our data give the evidence of a secondary field-induced multiferroic transition at T$_{{\rm N3}}$ (slightly lower than T$_{{\rm N2}}$).  Our speculation about the new field-induced multiferroic state requires the further confirmation from future  neutron diffraction experiments in strong magnetic fields.

\begin{acknowledgments}
We thank Dr  Ming-Jye Wang and Wei-Li Lee for experimental support and helpful discussions. Martin acknowledges the support from the Tech Trek Program of National Science Council of Taiwan.
\end{acknowledgments}


\begin{thebibliography}{}
\bibitem{TbMnO3}
T. Kimura, T. Goto, H. Shintani, K. Ishizaka, T. Arima, and Y. Tokura, Nature (London) {\bf 426}, 55 (2003).
\bibitem{TbMn2O5}
N. Hur, S. Park, P.A. Sharma, J.S. Ahn, S. Guha, and S-W. Cheong, Nature (London) {\bf 429}, 392 (2004).
\bibitem{Fiebig}
M. Fiebig, J. Phys. D {\bf 38}, R123 (2005).
\bibitem{CheongNM}
S.-W. Cheong and M. Mostovoy, Nat. Mater. {\bf 6}, 13 (2007).


\bibitem{LiuJM}
K.F. Wang, J.-M.  Liu, and Z.F. Ren, Adv. Phys.{\bf 58}, 321(2009).


\bibitem{DM}
T. Moriya, Phys. Rev. {\bf 120}, 91 (1960).
\bibitem{Dagotto}
I. A. Sergienko and E. Dagotto, Phys. Rev. B {\bf 73}, 094434 (2006).
\bibitem{Nagaosa}
H. Katsura, N. Nagaosa, and A. V. Balatsky, Phys. Rev. Lett. {\bf 95}, 057205 (2005).
\bibitem{Mostovoy}
M. Mostovoy, Phys. Rev. Lett. {\bf 96}, 067601 (2006).
\bibitem{Lawes}    	
G. Lawes, T. Kimura, C. M. Varma, M. A. Subramanian, N. Rogado, R.J. Cava and A.P. Ramirez, Prog. Solid. Stat. Chem, {\bf 37}, 40(2009)
\bibitem{1971}
L.M. Levinson, B.M. Wanklyn, J. Solid State Chem. {\bf 3}, 131 (1971).
\bibitem{1972}
B. Robertson, E. Kostiner, J. Solid State Chem. {\bf 4}, 29  (1972).
\bibitem{2008}
Z. He, J. Yamaura, and Y. Ueda, J. Solid State Chem. {\bf 181}, 2346 (2008).
\bibitem{Kundys1}
B. Kundys, C. Martin, and C. Simon, Phys. Rev. B {\bf 80}, 172103 (2009).
\bibitem{Kundys2}
A. Daoud-Aladine, B. Kundys, C. Martin, P. G. Radaelli, P. J. Brown, C. Simon, and L. C. Chapon, Phys. Rev. B {\bf 80},
220402(R)(2009).
\bibitem{Lawes1}
A. Dixit and G. Lawes, J. Phys.: Condens. Matter {\bf 21}, 456003 (2009).
\bibitem{Lawes2}
A. Dixit, G. Lawes, and A.B. Harris, Phys. Rev. B {\bf 82}, 024430(2010).
\end{thebibliography}
\end{document}